\documentclass[conference]{IEEEtran}
\IEEEoverridecommandlockouts
\usepackage{cite}
\usepackage{amsmath,amssymb,amsfonts}
\usepackage{algorithmic}
\usepackage{graphicx}
\usepackage{textcomp}
\usepackage{xcolor}
\def\BibTeX{{\rm B\kern-.05em{\sc i\kern-.025em b}\kern-.08em
    T\kern-.1667em\lower.7ex\hbox{E}\kern-.125emX}}
\begin{document}

\title{A novel Algorithm for Hydrostatic-mechanical Mobile Machines
with a Dual-Clutch Transmission \\
}

\makeatletter
\newcommand{\linebreakand}{%
  \end{@IEEEauthorhalign}
  \hfill\mbox{}\par
  \mbox{}\hfill\begin{@IEEEauthorhalign}
}

\makeatother

\author{
  \IEEEauthorblockN{1\textsuperscript{st} Yusheng Xiang}
  \IEEEauthorblockA{\textit{Institute of Vehicle System Technology} \\
    \textit{Karlsruhe Institute of Technology}\\
    Karlsruhe, Germany \\
    yusheng.xiang@partner.kit.edu}
  \and
  \IEEEauthorblockN{2\textsuperscript{nd} Ruoyu Li }
  \IEEEauthorblockA{\textit{Institute of Vehicle System Technology} \\
    \textit{Karlsruhe Institute of Technology}\\
    Karlsruhe, Germany \\
    ruoyu.li@kit.edu}
  \and
  \IEEEauthorblockN{3\textsuperscript{rd} Christine Brach}
      \IEEEauthorblockA{\textit{Department of Mobile Hydraulics} \\
    \textit{Robert Bosch GmbH}\\
    Elchingen, Germany \\
    christine.brach@boschrexroth.de}
  \linebreakand 
  \IEEEauthorblockN{4\textsuperscript{th} Xiaole Liu}
  \IEEEauthorblockA{\textit{Department of Electrical and Computer Engineering} \\
    \textit{Technical University of Munich}\\
    Munich, Germany \\
    xiaole.liu@tum.de}
  \and
  \IEEEauthorblockN{5\textsuperscript{th} Marcus Geimer}
  \IEEEauthorblockA{\textit{Institute of Vehicle System Technology} \\
    \textit{Karlsruhe Institute of Technology}\\
    Karlsruhe, Germany \\
    marcus.geimer@kit.edu}
}
\maketitle

\begin{abstract}
Mobile machines using a hydrostatic transmission is highly efficient under lower working-speed condition but less capable at higher transport velocities. To enhance overall efficiency, we have improved the powertrain design by combining a hydrostatic transmission with a dual-clutch transmission (DCT). Compared with other mechanical gearboxes, the DCT avoids the interruption of torque transmission in the process of shifting without sacrificing more transmission efficiency. However, there are some problems of unstable torque transmission during the shifting process, and an excessive torque drop occurring at the end of the gear shift, which result in a poor drive comfort. To enhance the performance of the novel structural possibility of powertrain design, we designed a novel control strategy for the motor torque and the clutch torques during the shifting process. The controller’s task is tracking the target drive torque and completing the gear shift as quickly as possible with acceptable smoothness requirements. In the process of the controller design, a specific control algorithm is firstly designed for a 10 ton wheel loader. The model-based simulation has validated the control effect. As a result, the control strategy employing clutch and motor torque control to achieve a smooth shifting process is feasible since drive torque is well tracked, and highly dynamical actuators are not required. Furthermore, only two calibration parameters are designed to adjust the control performance systematically. The novel control strategy is generalized for other mobile machines with different sizes.

\end{abstract}


\section{Introduction}

The combination of a hydrostatic transmission and a mechanical gearbox can improve not only the system efficiency but also the usability of mobile machines. The reason is the shiftable mechanical gearbox can make better working conditions for hydraulic components. Currently, a commonly used structure is the combination of a hydrostatic transmission (HST) with a summation gearbox, which has at least two hydraulic motors to propel the construction machinery. However, the additional cost of the second motor limits the scope of using of this structure in the market of medium-sized mobile machines. By contrast, we propose a novel structure that combines the HST with a dual-clutch transmission, see Fig. \ref{fig:Dualclutchpowertrain}. With this novel combination, we  only need one hydraulic motor and thus reduce the total transmission costs compared with the one with a summation gearbox. However, such a drivetrain solution did not have a considerable
market share in a long time. One of the reasons is the fluctuation of output torque during the shifting process. Based on the primary torque control concept from Bosch Rexroth \cite{Mutschler.2018,Xiang.2020,Xiang1.2020}, we developed a controller to optimize the shifting process of a dual-clutch transmission resulting in a fast and smooth shifting process. As a basic idea of this algorithm, the hydraulic motor and the clutches should cooperate with each other to accurately track the desired drive torque. 
In this paper, we will first reveal the mechanism of the problem mentioned above and then introduce the algorithm which can optimize the performance of the HST+DCT drivetrain. In addition, the algorithm has an easy calibration process since only two calibration parameters are needed. 
\begin{figure}[htbp]
\centering
\includegraphics[width=3in]{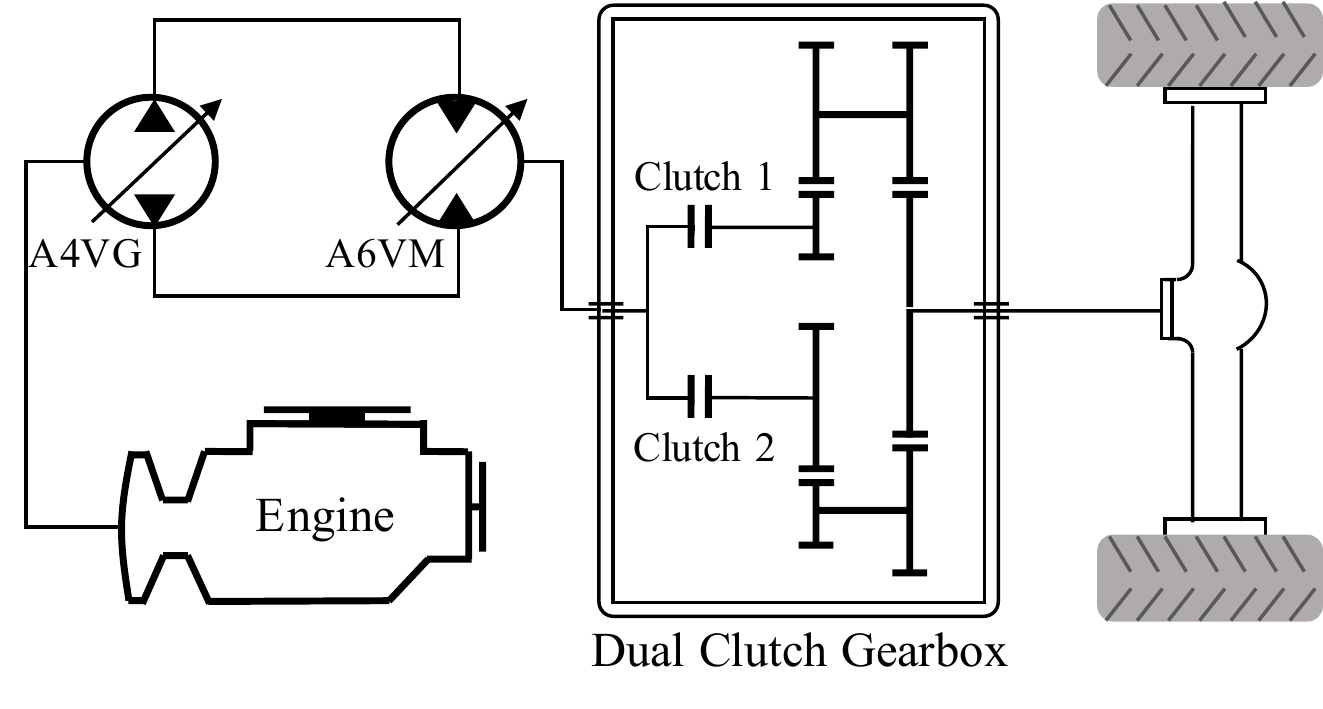}
\caption{HST+DCT drivetrain}
\label{fig:Dualclutchpowertrain}
\end{figure}

\section{Related works}

Currently, there are a series of researches about the control of dual clutches gearbox in the automobile industry. Fischer demonstrates the typical problem by using a dual-clutch gearbox \cite{Fischer.2015}. While upshifting, the torque goes firstly down and then goes up. At the end of the shifting process, there is a sudden torque drop, resulting in a vehicle jerk \cite{Fischer.2015}. Oh proposes an original approach for the driveline modeling so that a complex clutch friction model can be avoided \cite{Oh.2014}. Moreover, many researchers propose their methods to improve the shift quality, i.e., short shift duration and smooth shift process \cite{Walker.2011, Song.2005, vanBerkel.2014, Kulkarni.2007, Liu.2014, Kim.2017}. Van Berkel introduces the concept of clutch-engagement phases so that a smooth clutch engagement to track the demanded torque without a noticeable torque dip precisely is accomplished \cite{vanBerkel.2014}. Kim develops an effective upper-level controller based on the optimal control allocation to product the most suitable torque trajectories of the engine and clutches \cite{Kim.2017}.

In the fields of mobile machines, the drivetrains using clutches to output continuous drive torque are called power shift transmission since they might have more than two clutches \cite{Geimer.2014}. Bugusch shows that a powershift-hydrostatical drivetrain with three gears, developed by Göllner \cite{Gollner.2007}, can improve the vehicle efficiency by selecting the gear appropriately \cite{Bagusch.}. Mörsch has pointed out in a patent to improve the drive comfort during the shifting process, at least one variable should be used \cite{Moersch.2019}.

\section{Problem statement}

According to the rigorous literature review, we can make some conclusions from the previous studies. 
Firstly, most researches try to increase the number of gears so that they can decrease the transmission ratio difference between each gear, which undoubtedly relieves the difficulties of the control algorithm. Secondly, for the cost reason, precise measurement of the signal is still challenging, and thus the actuator dynamics should be taken into account while designing the controller. Last but not least, although many companies in the field of mobile machines show great interest in using power shift transmission, they did not propose a concrete algorithm to achieve good shift quality. Consequently, the drive comfort during the shifting process of the HST+DCT drivetrain was never satisfying, and no wonder it has only a small market share. We conjecture it might be that it is impossible to adopt the easiest method that increases the number of mechanical gears on a hydrostatic mobile machine.  

\section{Goal of this research}
The goal of our study is to find out a suitable control concept which guarantees a fast and smooth shift process of HST+DCT drivetrain for the whole shifting process. Concretely, we control the actual tractive effort so that it is almost the same as the desired tractive effort during the shifting process, as \eqref{eq:goal}
\begin{equation}
\int_{t_s}^{t_e} [T_{act}(t) - T_{des}(t)]^2 dt \rightarrow 0\label{eq:goal}
\end{equation}
where $t_s$ and $t_e$ denote the shifting start and end time. 

\section{System modelling}

\subsection{Powertrain}

In order to obtain more insight into the control algorithm, we greatly simply the dual-clutches model according to \cite{Fischer.2015}. The dual-clutch gearbox has primary parts (subsystem 1) and secondary parts (subsystem 2), corresponding to the part before the reduction gears and the part after the reduction gears, separately. Fig. \ref{fig:powertrain} illustrates the simplified powertrain model we use.

\begin{figure}[htbp]
\includegraphics[width=3.5in]{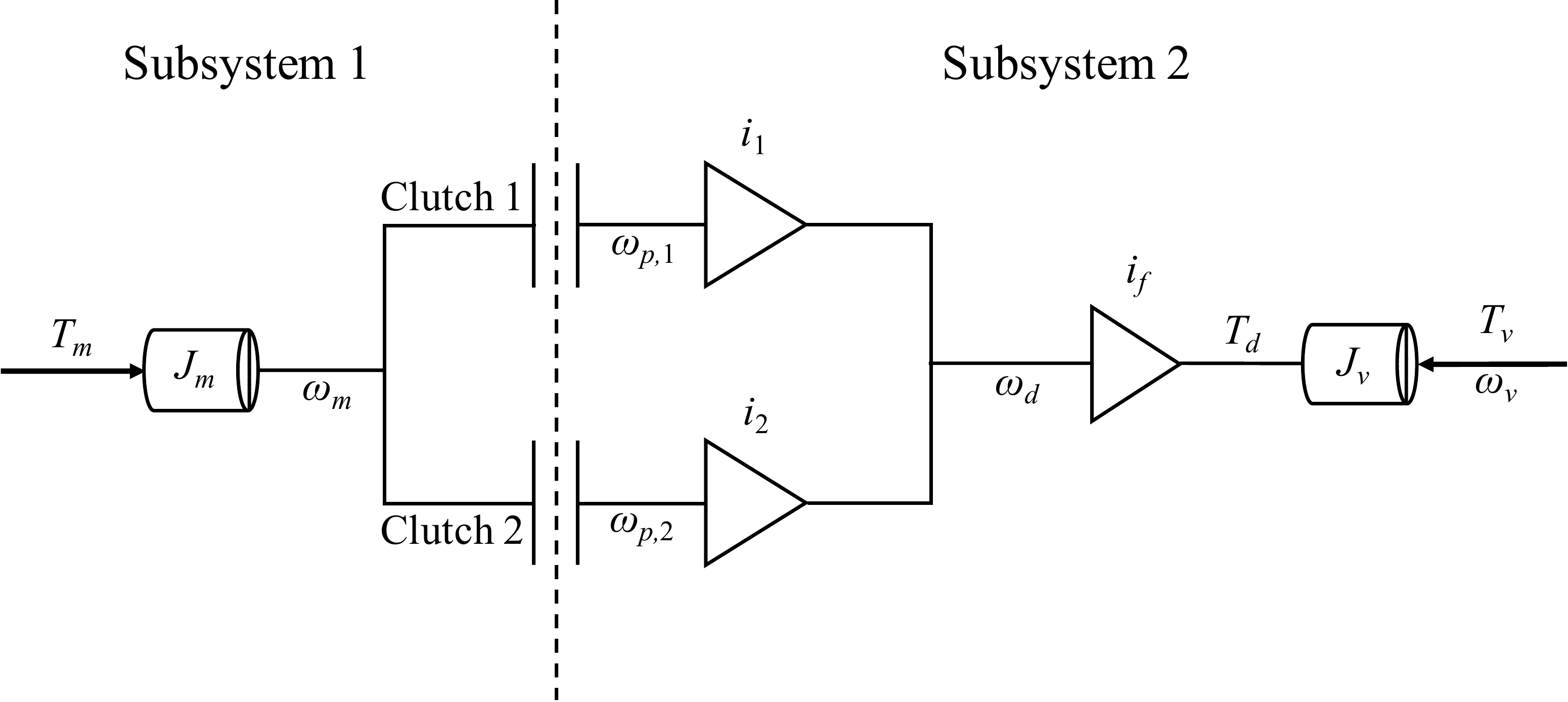}
\caption{Simplified powertrain model: $\omega_{p,1}$ is the rotation speed of primary part of the first gear, $T_m$ in the output torque of hydraulic motor, $T_d$ is the drive torque on the wheels, and $T_v$ is the resistant force.}
\label{fig:powertrain}
\end{figure}

The driving torque $T_d$ in the transmission chain is completely transmitted by the torques on both clutches $T_{c1}$ and $T_{c2}$. Moreover, the transmission loss is given by the final drive, while the damping and stiffness of the drive shaft are not taken into account.

The power train dynamic is described as follows; if both of the two clutches are not in an engagement state, the transmission chain is a 2DOF system. No matter upshifting or downshifting, we produce a positive drive torque on the wheels. Thus, the 2DOF system can be drawn as 

\begin{eqnarray}
\left\{
\begin{array}{lr}
J_m \cdot \dot\omega_m = T_m - T_{c1} - T_{c2} \label{cp4_dwe} \\
J_v \cdot \dot\omega_v = T_d - T_v \label{cp4_dwv}
\end{array}
\right.
\end{eqnarray}
where the value of driving torque depends on the values of the two clutch torques and the transmission parameters:
\begin{eqnarray}
T_d &=& ( T_{c1} \cdot i_1 + T_{c2} \cdot i_2 ) \cdot i_{\text{final}} \cdot \eta \label{cp4_td} \\
T_{c1} &=& \mu \cdot F_{N,c1} \label{cp4_tc1} \\
T_{c2} &=& \mu \cdot F_{N,c2} \label{cp4_tc2}
\end{eqnarray}

When the vehicle runs stably in gear 1, i.e., the clutch 1 is in engagement, the degree of freedom of the system is reduced by 1, because the rotational speed of motor shaft, the secondary side of the first clutch,  and final-drive shaft must satisfy the following kinematics constraints.
\begin{eqnarray}
\left\{
\begin{array}{lr}
\omega_m = \omega_L \\
\dot\omega_m = \dot\omega_L \\
\dot\omega_L = \dot\omega_v \cdot i_{\text{final}} \cdot i_1
\end{array}
\right.
\end{eqnarray}
Although the final-drive shaft rotates at an unidentical speed, the ratio is fixed. In order to simplify the expression, the other clutch is set to be entirely disengaged without sliding, $T_{c2} = 0$; thus, system equation is rewritten to
\begin{equation}
(J_v + J_m \cdot i_1^{2} \cdot i_{\text{final}}^{2} \cdot \eta ) \cdot \dot\omega_v = T_m \cdot i_1 \cdot i_{\text{final}} \cdot \eta - T_v 
\end{equation}
and the torque on the first clutch is calculated as follows:

\begin{equation}
T_{c1,d} = \frac{T_m \cdot J_v + J_m \cdot T_v \cdot i_{\text{final}} \cdot i_1}{J_v + i_1^{2} \cdot i_{\text{final}}^{2} \cdot \eta \cdot J_m} \label{cp4_tc1_engage}
\end{equation}

If the first clutch is in the engagement state, its torque is no longer controlled only by the pressure acting on it. The result obtained by \eqref{cp4_tc1_engage} is the clutch torque required to meet the dynamics and kinematic constraints. The pressure on the clutch friction plate represents the maximum torque which the clutch can transmit.
\begin{equation}
\mu \cdot F_{N,c1} = T_{c1,{\text{Max}}}
\end{equation}
Only when the following conditions are met, the clutch maintains the engagement state; otherwise, it will start to slide.\begin{equation}
T_{c{\text{Max}}} \ge T_{c1,d}
\end{equation}

Normally, the driving resistance of the vehicle is more difficult to measure or estimate. From another perspective, the desired clutch torque can be calculated from the torque balance on the motor side.
\begin{eqnarray}
(T_m - T_{c1,d}) &=& J_m \cdot \dot\omega_m = J_m \cdot \dot\omega_v \cdot i_{\text{final}} \cdot i_1 \\
T_{c1,d} &=& T_m - J_m \cdot \dot\omega_v \cdot i_{\text{final}} \cdot i_1 \label{cp4_tc1_desired}
\end{eqnarray}

It can be seen from \eqref{cp4_tc1_desired}, in the case where the motor output torque and the rotational acceleration of the drive shaft are known, the value of the clutch torque required to maintain the stable running of the vehicle in gear one can be calculated, wherein the first derivative of vehicle speed can determine the rotational acceleration of the drive shaft.

It can be concluded that because the moment of inertia of the motor torque exists, the torque acting on the clutch is unidentical to the torque output by the motor, i.e., the driving torque acting on the transmission shaft cannot be directly converted from the motor output torque. It is also necessary to consider the torque required to accelerate the motor's moment of inertia.

Similarly, when the vehicle is running stably at gear 2, the equation of 1 DOF system is given without derivation as: 
the 1 DOF system equation for gear 2
\begin{equation}
(J_v + J_m \cdot i_2^{2} \cdot i_{\text{final}}^{2} \cdot \eta ) \cdot \dot\omega_v = T_m \cdot i_2 \cdot i_{\text{final}} \cdot \eta - T_v 
\end{equation}

The torque required to ensure that the clutch 2 does not slip. 
\begin{equation}
T_{c2,d} = T_m - J_m \cdot \dot\omega_v \cdot i_{\text{final}} \cdot i_2
\end{equation}

\subsection{Actuator}

The input signal of the actuator is the output of the controller. Although the controller's instructions are correct, the system may not perform well due to the slow response of the actuators. Thus, the controller must output the command in which the dynamics behavior of actuators is already considered. The dynamic response of the actuator is modeled as 1st order differential equation \cite{vanBerkel.2014}, 
\begin{eqnarray}
\dot{T}_{c1}(t) &=& \frac{1}{\theta_{c1}}(T_{c1}^{'}(t)-T_{c1}(t))\\
\dot{T}_{c2}(t) &=& \frac{1}{\theta_{c2}}(T_{c2}^{'}(t)-T_{c2}(t))\\
\dot{T}_m(t) &=& \frac{1}{\theta_m}(T_m^{'}(t)-T_m(t))
\end{eqnarray}
$\theta_m$ and $\theta_c$ are the actuators' time constant.  

\section{Power shift Controller}

The power shift controller consists of two parts, the Phase Selector and the Torque Generator. According to different shifting requirements (upshift or downshift), the phase selector selects the corresponding control mode. Also, the shifting process is further subdivided into different stages. The generator outputs a specific control signal according to the phase signal given by the phase selector.

\subsection{Control Schema}
The control framework based on the torque based control consists of a target torque generator, a powershift controller, an actuator dynamic module, and a power train model. The schematic diagram is shown in Fig. \ref{cp4_1fig}.
\begin{figure}[htbp]
	\centering
	\includegraphics[width=3.5in]{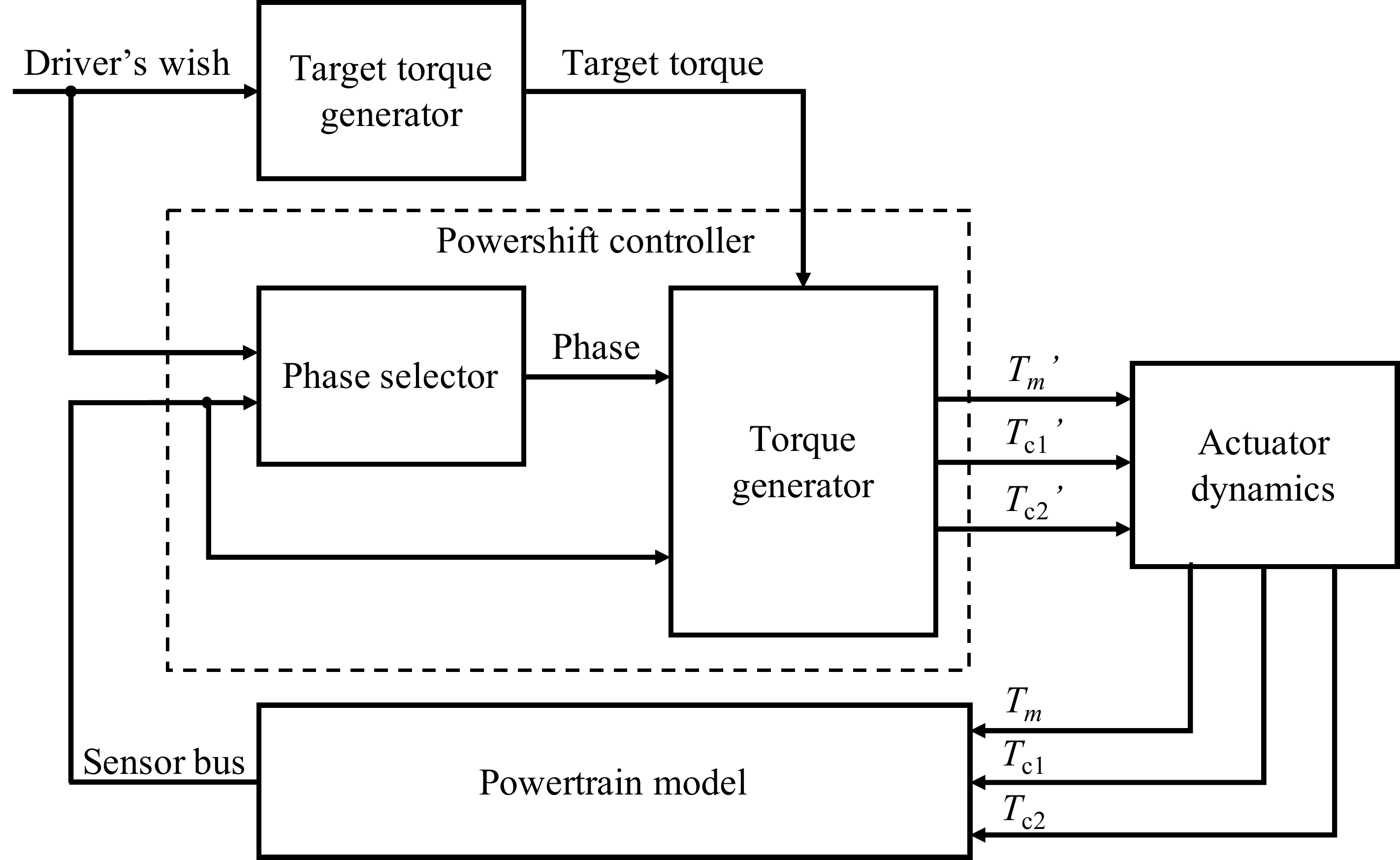}
	\caption{Power shift controller framework}
	\label{cp4_1fig}
\end{figure}



The controller outputs to the vehicle model three torque variables that affect the shifting process: motor torque, higher gear clutch torque, and lower gear clutch torque. The controller determines, according to the target, the driving condition, the driver's wish, and the state variable of the vehicle as a boundary condition, whether the vehicle can complete the given shift command in the current driving situation, and if so, outputs the corresponding control variable, if not, outputs the control variable of the closest target requirement.

For the torque control of the shifting process, it is necessary to pre-consider the blockage of the actuator when calculating the control variable to eliminate the influence of the dynamic response of the actuator on the control performance in the actual process. 


\subsection{Phase Selector}

Fig. \ref{cp4_4fig} shows the schematic diagram of phase selector. The shifting process is divided into several sub-stages to distinguish different requirements of the power shift controller. Drive mode represents the drive chain, which is in the process of upshifting, downshifting, or maintaining the current gear. In the specific algorithm implementation, the upshift is divided into torque phase and inertia phase. Moreover, the inertia phase is subdivided into fast reduction of slip speed, controlled reduction of slip speed, and forcibly engaged phase.
\begin{figure}[htbp]
	\centering
	\includegraphics[width=3.5in]{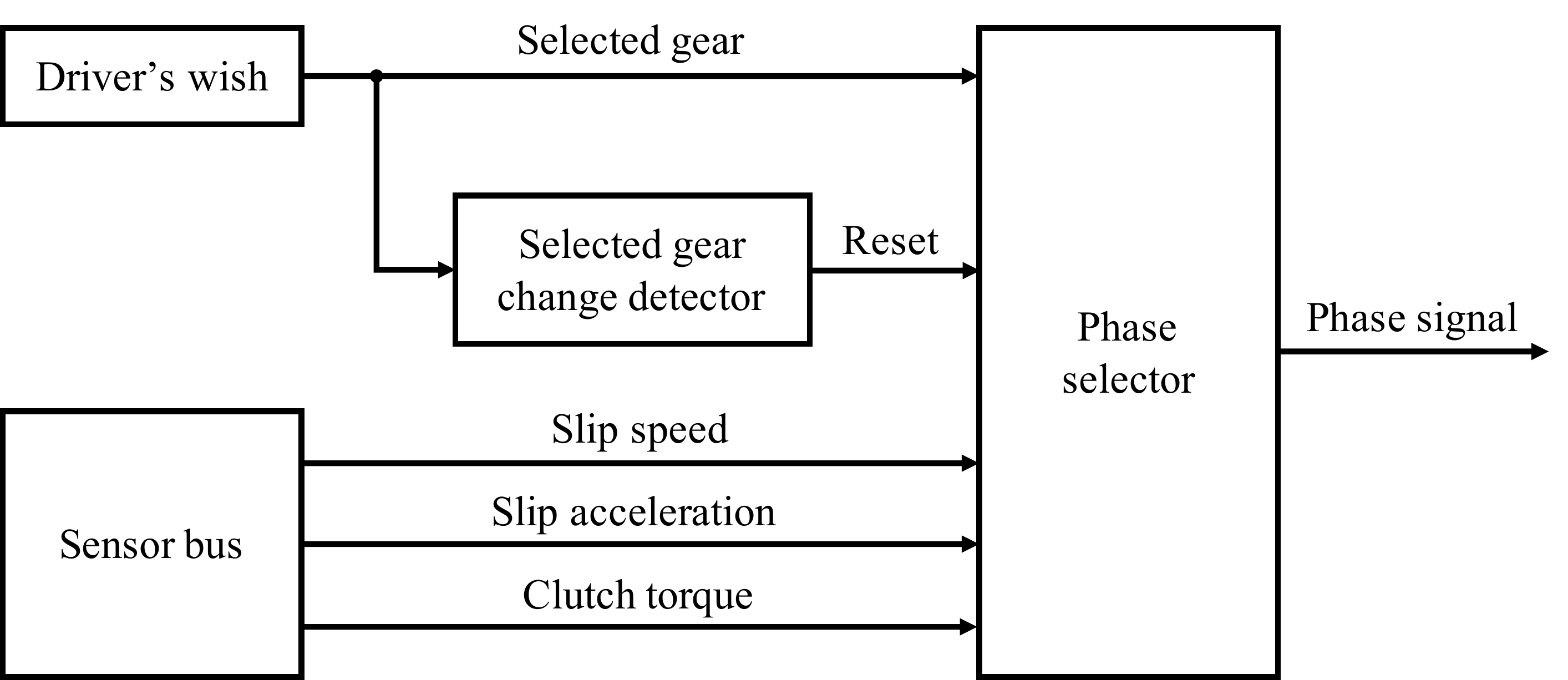}
	\caption{Phase selector}
	\label{cp4_4fig}
\end{figure}
 
In the downshift process,  there is no apparent torque phase. The cross-shift occurs during the second inertia phase. Concretely, it begins at the time when the $\omega_s$ first reaches the $\Omega_2$.  
 
The phase transition is determined according to the vehicle state. For example, in the torque phase, when the phase selector judges whether the cross shift has ended, it is necessary to pass the clutch torque feedback to indicate whether the clutch has been disengaged. In the inertia phase, it determines whether the controller flows into the next stage base on the slip speed and slip acceleration.

\subsection{Torque Generator}

The concept is shown in Fig. \ref{cp4_5fig}, the torque generator is the core module of the torque controller. The controller's input signals are $T_{d,target}$, drive mode, phase signal, and vehicle state. The tracking of target driving torque is the main task of the power shift controller, drive mode, and phase can be seen as boundary conditions, and vehicle state is the feedback from power train dynamics.

The clutch torque generator needs to give the clutch torque and its rate of change required to track the target driving torque according to the boundary conditions. Furthermore, the motor torque generator needs to calculate the motor torque according to the vehicle state feedback to meet the controllable condition of the clutch torque.

If the torque balance is destroyed during the shifting process, a completely different control performance will be produced; thus, a precise torque control is required. It is necessary to pre-consider the effect of the actuator delay, i.e. the inversed actuator dynamics will be taken into account when designing the control signal. In this way, after the actuator delay the control signal can still complete the control tasks.
\begin{figure}[htbp]
	\centering
	\includegraphics[width=3.5in]{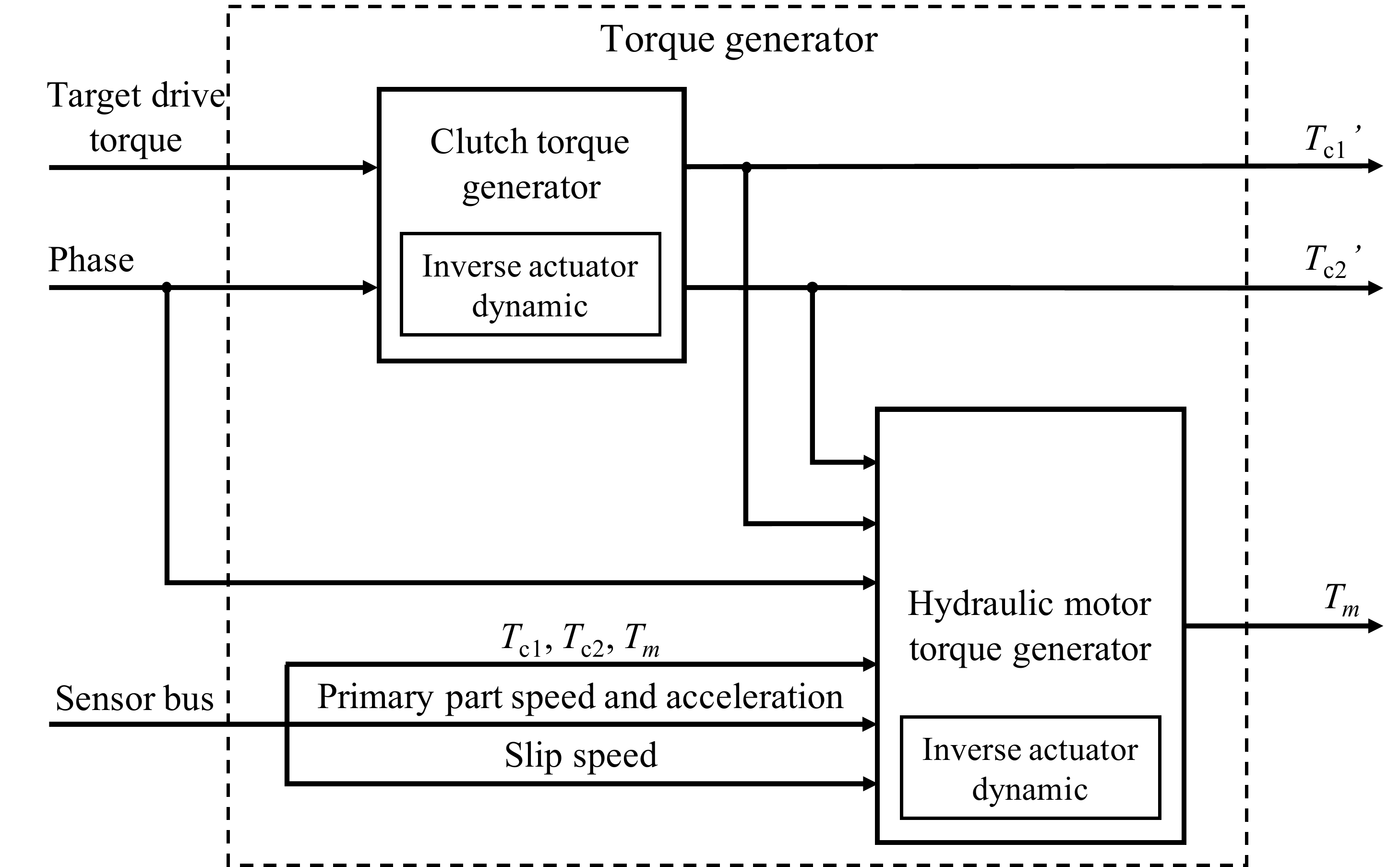}
	\caption{Torque generator}
	\label{cp4_5fig}
\end{figure}

\section{Control algorithm}

\subsection{Torque phase}

In the upshifting process, the first torque drop happens in the torque phase \cite{Goetz.2003} when the cross shift occurs.  The core idea of the solution is to let the hydraulic motor rotate faster than the primary part of dual-clutch transmission so that the system is decoupled and can be controlled separately. In the case of upshifting, at the initial point of the torque phase, there is no slip, i.e., $ \omega_s=0 $. The concrete control idea can be illustrated with the following equations.
\begin{eqnarray}
T_d &=& T_{c1} \cdot i_1 + T_{c2} \cdot i_2  \label{eq:driveforce}\\
T_{m} &=& T_{c1} + T_{c2} + J_m \cdot \dot\omega_m \label{eq:motorforce}\\
\dot\omega_s&=& \dot\omega_m - \dot\omega_p > 0 \label{eq:condiction} 
\end{eqnarray}

The eq. \eqref{eq:condiction} shows the necessary condition to achieve the control algorithm. Otherwise, the eq. \eqref{eq:driveforce} and eq. \eqref{eq:motorforce} will have a different sign, which makes the control dramatically difficult. According to eq. \eqref{eq:driveforce}, the transmitted torque of the second clutch $ T_{c2}$ can be calculated based on $ T_{d}$ and $ T_{c1}$. Since $ T_{d}$ goes to zero while upshifting with a specific slope given by the designer, we can acquire $ T_{c2}$. After we obtain $ T_{c2}$, the desired hydraulic motor torque can be acquired. For the purpose of the robustness of the controller, the $\ddot\omega_{s,set}$ should be selected carefully to guarantee the slip $ \omega_s $  is always positive. However, it might lead to an unnecessary additional friction loss of the clutches. Therefore, this safety value should be as small as possible from the respective of energy loss. Taking the actuator dynamics into account, we describe the control signal of hydrostatic motor torque as eq. \eqref{eq:torqueDesired},
\begin{equation}
\begin{aligned}
T_m^{'} = &T_m + \frac{\theta_m}{\theta_c} \cdot \left( T_{c1}^{'} - T_c \right) + \frac{\theta_m}{\theta_c} \cdot \left( T_{c2}^{'} - T_{c2} \right) \\
&+ \theta_m \cdot J_m \cdot i_{\text{final}} \cdot i_{\text{current}} \cdot \ddot\omega_v + \ddot\omega_{s,set} \cdot \theta_m \cdot J_m \label{eq:torqueDesired}
\end{aligned}
\end{equation}

\subsection{Inertia phase}

In the inertia phase of the upshifting process, the hydraulic motor rotation speed must slow down to the corresponding upcoming gear. In this case, the slip between hydraulic motor and primary part of the dual-clutch absolutely exists. Thus, the output torque on vehicle wheels is further controlled by clutches. However, the challenge of implementing a control algorithm on a vehicle is that the transmit torque when the clutch is just engaged. At this time, the system will lose one degree of freedom, i.e., from 2 DoF to 1 DoF system resulting in the last torque dip. Thus, the driving force might change immediately. The following equations illustrate the solution to this problem.

The drive torque before engagement $T_{c,tar}(t^{-})$ is demonstrated as \eqref{eq:BeforeEngagegd},
\begin{equation}
\begin{aligned}
T_{c,tar}(t^{-}) = &\frac{1}{J_v + J_m \cdot i_{tar}^{2} \cdot i_{\text{final}}^{2} \cdot \eta} \cdot (J_v \cdot T_m(t^{-}) +\\
&J_m \cdot T_v(t^{-}) \cdot i_{tar} \cdot i_{\text{final}} - \dot\omega_{s,tar}(t^{-}) \cdot J_m \cdot J_v) \label{eq:BeforeEngagegd}
\end{aligned}
\end{equation}
In contrast, the drive torque after engagement $T_{c,tar}(t^{+})$ is demonstrated as \eqref{eq:BeforeEngagegd},
\begin{equation}
T_{c,tar}(t^{+}) = \frac{J_v \cdot T_m(t^{-}) + J_m \cdot T_v(t^{-}) \cdot i_{tar} \cdot i_{\text{final}}}{J_v + J_m \cdot i_{tar}^{2} \cdot i_{\text{final}}^{2} \cdot \eta} \label{eq:AfterEngagegd}
\end{equation}
Since the motor torque and the driving resistance is continuous,
\begin{eqnarray}
T_m(t^{+}) &=& T_m(t^{-})\\
T_v(t^{+}) &=& T_v(t^{-})
\end{eqnarray} 
The drive torque, i.e., the difference between the drive chain output before$T_d(t^{-})$ and after $T_d(t^{+})$ the engagement are the object of the controller to minimize
\begin{equation}
\begin{aligned}
T_d(t^{+}) - T_d(t^{-}) =& T_{c,tar}(t^{+}) \cdot i_{tar} \cdot i_{final} \cdot \eta  \\
                         & - T_{c,tar}(t^{-}) \cdot i_{tar} \cdot i_{final} \cdot \eta  \label{eq:torquedifference}
\end{aligned}
\end{equation}
After $T_{c,tar}(t^{+})$ is substituted into the \eqref{eq:torquedifference}, the torque drop is interpreted as
\begin{equation}
T_d(t^{+}) - T_d(t^{-}) = \frac{i_{tar} \cdot i_{final} \cdot \eta}{J_v + J_m \cdot i_{tar}^{2} \cdot i_{final}^{2} \cdot \eta} \cdot \dot\omega_{s,tar}(t^{-}) \cdot J_m \cdot J_v \label{eq:TorqueDrop}
\end{equation}

As shown in \eqref{eq:TorqueDrop},  $\dot{\omega}_s$ is the only controllable variable since the other parameters are fixed after the vehicle is designed. Formally, if $\dot{\omega}_s$ is 0 when the clutch is just engaged, theoretically, there should not be a torque drop. However, to control the $\dot{\omega}_s$ is a difficult task in practice. First and foremost, $\dot{\omega}_s$ is challenging to measure because there is usually no acceleration sensor on the shaft to measure the rotation acceleration. Besides, to measure the slip acceleration, at least two acceleration sensors must be used. Moreover, the dynamics of actuators might be quite different and thus making the calibration process considerable effort. Last but not least, the acceleration sensors have higher noise in general and surely have a negative influence on control performance. Alternatively, we use an algorithm based on the concept from van Berkel \cite{vanBerkel.2014}, which only uses $\omega_s$ as the measured term. The figure illustrates the control algorithm in the inertia process for a downshift process since downshift is the more complicated process. In order to achieve the fast and smooth engagement, the inertia phase is further divided into five sub-phase: the first fast phase (P0), the first smooth phase (P1), the second fast phase (P2), the second smooth phase (P3), and the forcibly engaged phase (P4).

\begin{figure}[htbp]
	\centering
	\includegraphics[width=3.5in]{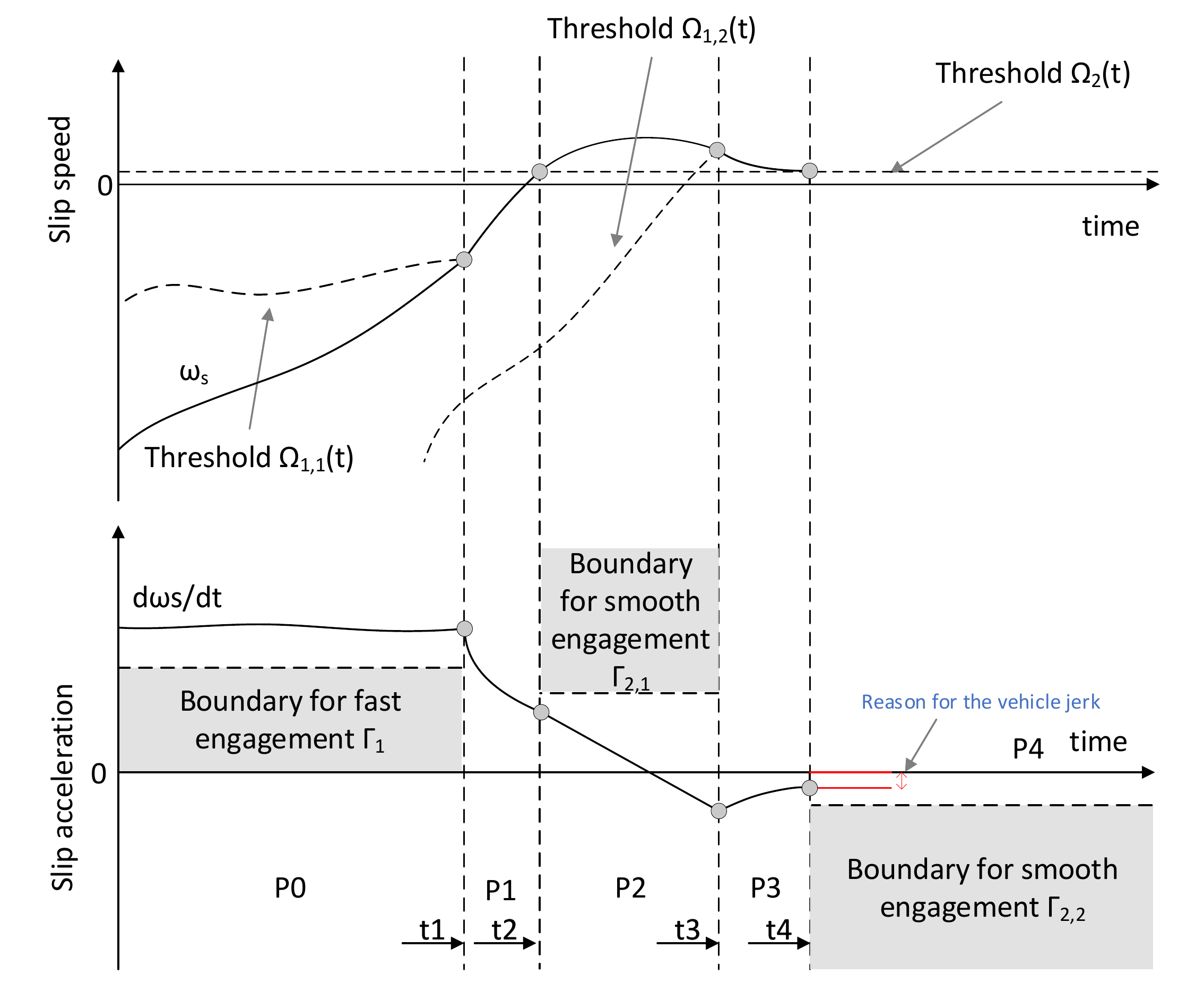}
	\caption{Algorithm explanation during the downshifting inertia phase: $\Omega_{1,1}(t)$ and $\Omega_{1,2}(t)$ are the triggers for the first and second smooth phase separately}
	\label{fig:Downshift_InertiaPhase}
\end{figure}

By deriving the differential equation containing the controlled object $\dot\omega_s$, the design method of the control variable could be obtained.
\begin{equation}
\ddot\omega_s = \frac{-1}{\theta_m} \cdot \dot\omega_s + \frac{1}{J_m} \cdot \left( \frac{1}{\theta_m} \cdot (T_m^{'} - T_c -J_m \cdot \dot\omega_p) - \dot{T}_c \right) - \ddot\omega_p \label{eq:ddws}
\end{equation}
When the form of \eqref{eq:ddws} could be converted into 
\begin{equation}
\ddot\omega_s + 2D \cdot \omega_n \cdot \dot\omega_s + \omega_n^2 \cdot  \omega_s = 0 \label{chap6_damp}
\end{equation}
under the condition of $D = 1$, this equation can be described as a critical damping system. The dynamics of slip speed is desired to be critical damped for the response would be fastest without overshooting which fitted the requirement of faster shifting process.

Here we introduce a constant offset $\Omega_0$, the modified slip speed is $x = \omega_s - \Omega_0 $. The value of $\Omega_0$ is peripheral since it will be eliminated in the transformation of equations. The equation after linearisation is, 
\begin{equation}
\ddot{x} + \frac{1}{\theta_m} \cdot \dot{x} - u = 0 \label{eq: u}
\end{equation}

Applying the condition of critically damped dynamic system the control variable $u$ is derived by comparing the coefficient of \eqref{eq: u} and \eqref{chap6_damp}
\begin{equation}
u = -\frac{1}{4\theta_m^{2}} \cdot x 
\end{equation}
with the condition $x(t_2)= \Omega(t_2)  - \Omega_0 = \Omega_2 - \Omega_0 $, $\dot{x}(t_1)= \Gamma_1$, and  $\dot{x}(t_2)= \Gamma_2$ are known, we can calculate  $x(t_1)= \Omega_1(t_1)$, which is the trigger for the fast phases.
\begin{eqnarray}
\Omega_1(t) &=& \Omega_2 + 2\theta_m \cdot (\Gamma_2 - \dot\omega_s(t))
\end{eqnarray}
\begin{equation}
u = - \frac{1}{4\theta_m^{2}} \cdot (\omega_s - \Omega_2 + \frac{\Gamma_2}{\lambda_m})
\end{equation}

\section{Simulation results}

In this section, the performance of the control algorithm would be validated. The drivetrain parameters used in simulations are based on a real wheel loader and shown in Tab. \ref{vehicleparameter}.

Notice that, for validation the algorithm, we consciously to let the drivetrain track a constant drive force demand since it must be the most challenging use case. 

As shown in Fig. \ref{fig:c7f18}, transmittable clutch torque controls the torque on the wheels during the whole downshifting process, while the hydraulic motor takes the responsibility to minimize the friction loss, positive power flow, and the fast and smooth engagement process. The next figure, Fig. \ref{fig:c7f19}, shows how the variables vary during the shifting process, where $ \omega_{s,1} $ denotes the slip speed between the hydraulic motor and primary first gear part of dual-clutch transmission $ \omega_{s,2} $ indicates the second gear. The torque drop by engaging is controllable and defined by $ \Gamma_{2,2} $ since the $ \omega_s = \Gamma_{2,2} $ at the moment of the engagement.

\begin{figure}[htbp]
	\centering
	\includegraphics[width=3.5in]{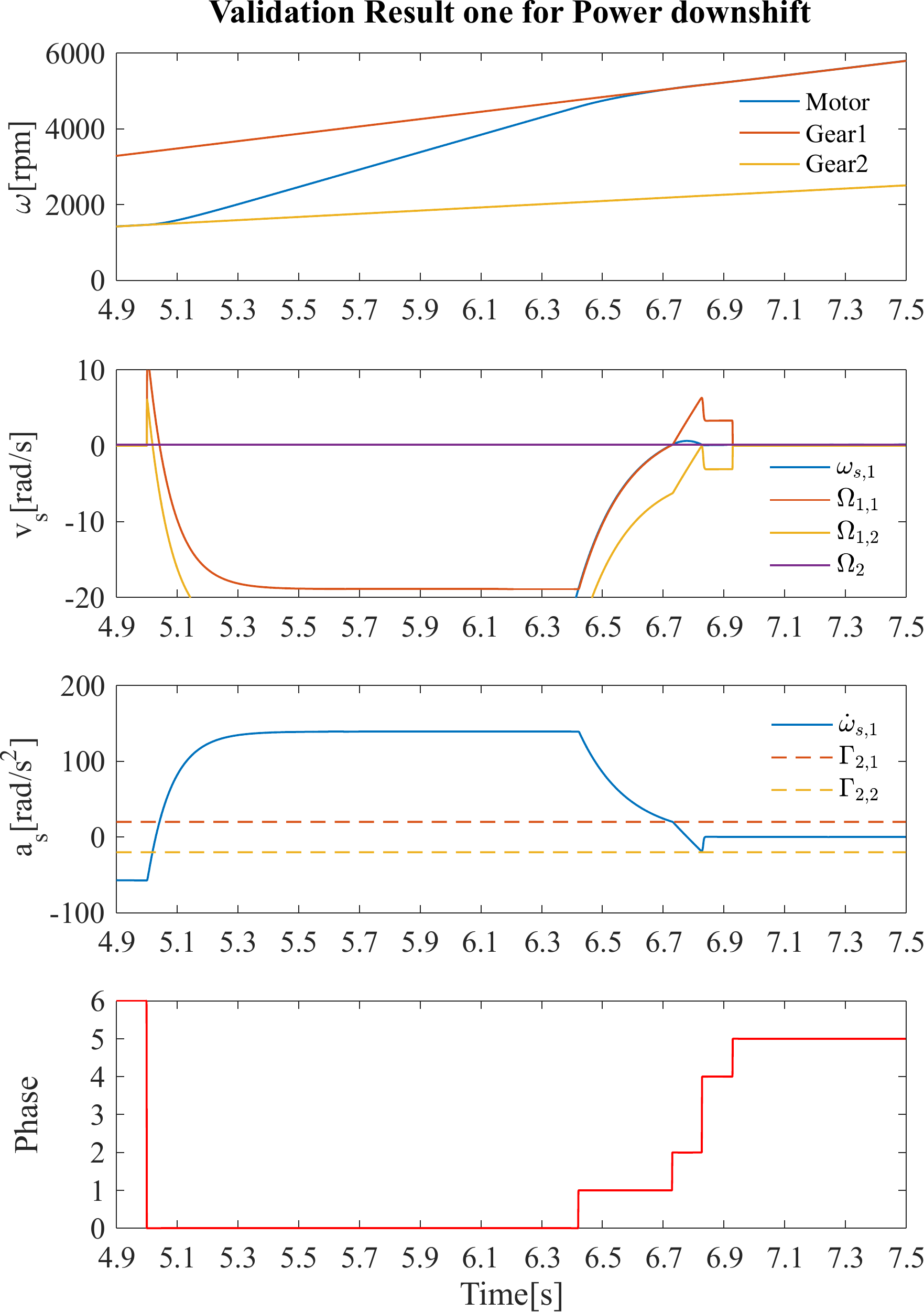}
	\caption{Simulation results illustrating the control strategy and the performance during the downshifting process}
	\label{fig:c7f18}
\end{figure}

\begin{figure}[htbp]
	\centering
	\includegraphics[width=3.5in]{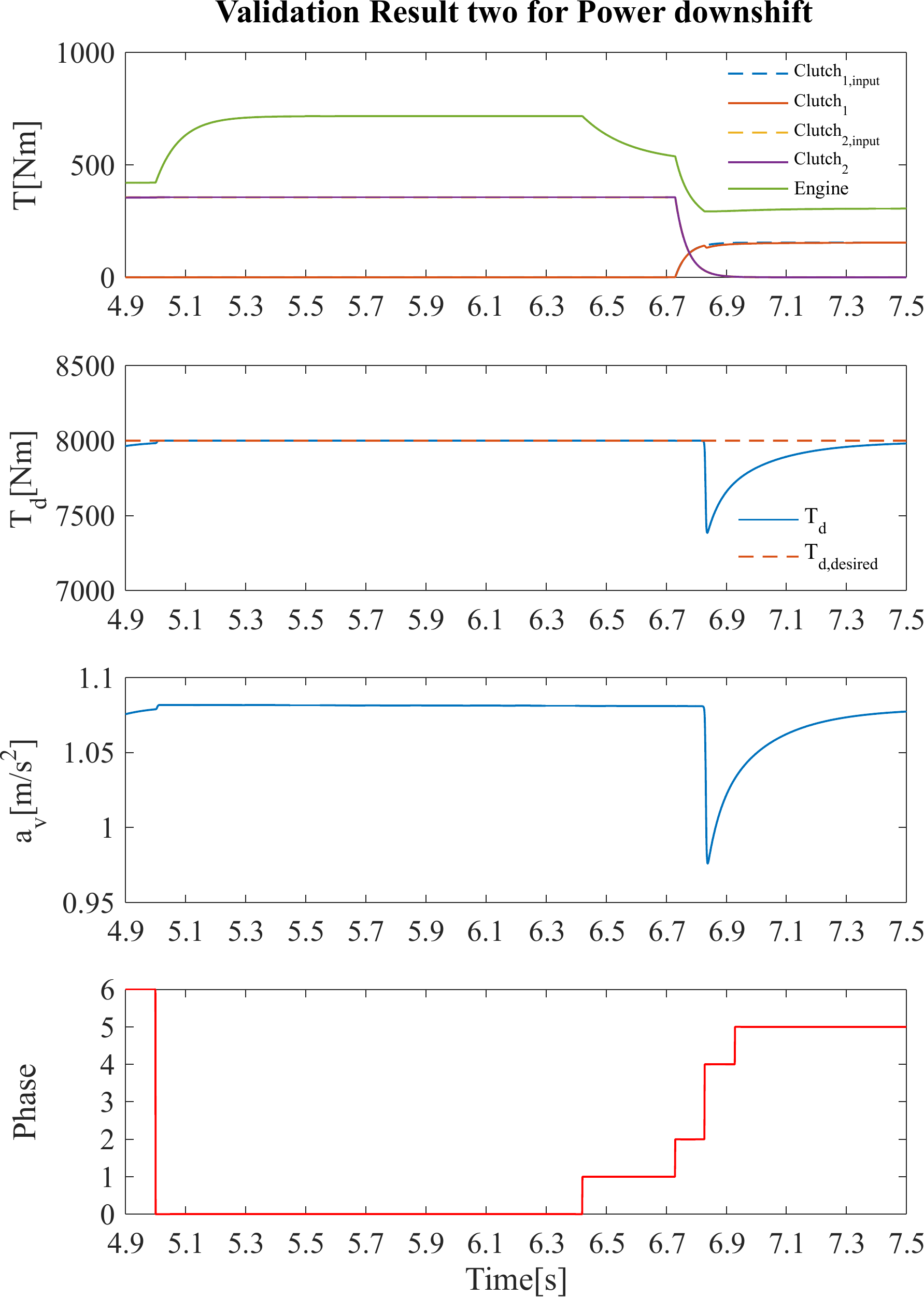}
	\caption{Control variables during the downshifting process}
	\label{fig:c7f19}
\end{figure}

As aforementioned, while the mechanism of the inertia phase before the final engagement is the same between upshifting and downshifting, the torque phase is the unique phase during the upshifting process. As shown in Fig. \ref{fig:upshift_1} and Fig. \ref{fig:upshift_2}, after the driver gives the upshift command,  the motor shaft speed $ \omega_m $  is slightly increased under motor torque control.  It can also be observed in the figures that by increasing the slip acceleration, the slip speed is controlled to be a small positive value. By ensuring that the clutch one is sliding, the condition that the clutch torque is controllable in the torque phase is satisfied.  

As the simulation results shown, the slip is the prerequisite of this algorithm and the correctly control of $\omega_s$ is  the core to achieve a fast and smooth shifting process. Also, there is no sudden change of control signal, which demonstrate that high dynamic actuators are not in need.

\section{Shift management}

Although the powershift with acceptable comfort is viable, the friction in the clutches is unavoidable. In order to extend the service time of the transmission, the powershift shall only occur when the vehicle is uphill and at a low velocity to avoid the roll down situation. In other scenarios, the interruption of tractive effort is not a problem so that we recommend to engage the upcoming gear after the ongoing gear was totally disengaged.  

\begin{figure}[htbp]
	\centering
	\includegraphics[width=3.5in]{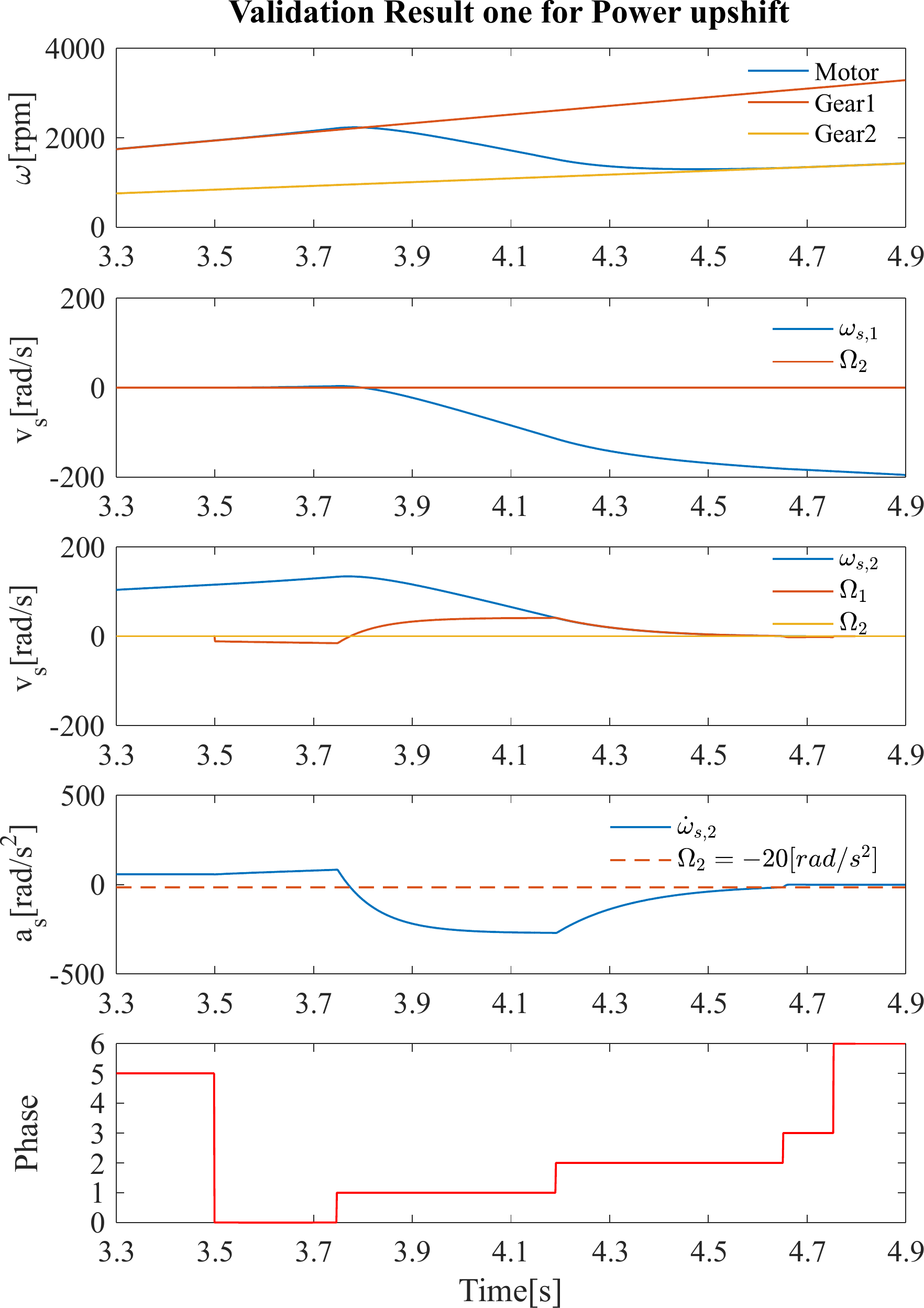}
	\caption{Simulation results illustrating the control strategy and the performance during the upshifting process}
	\label{fig:upshift_1}
\end{figure}

\begin{figure}[htbp]
	\centering
	\includegraphics[width=3.5in]{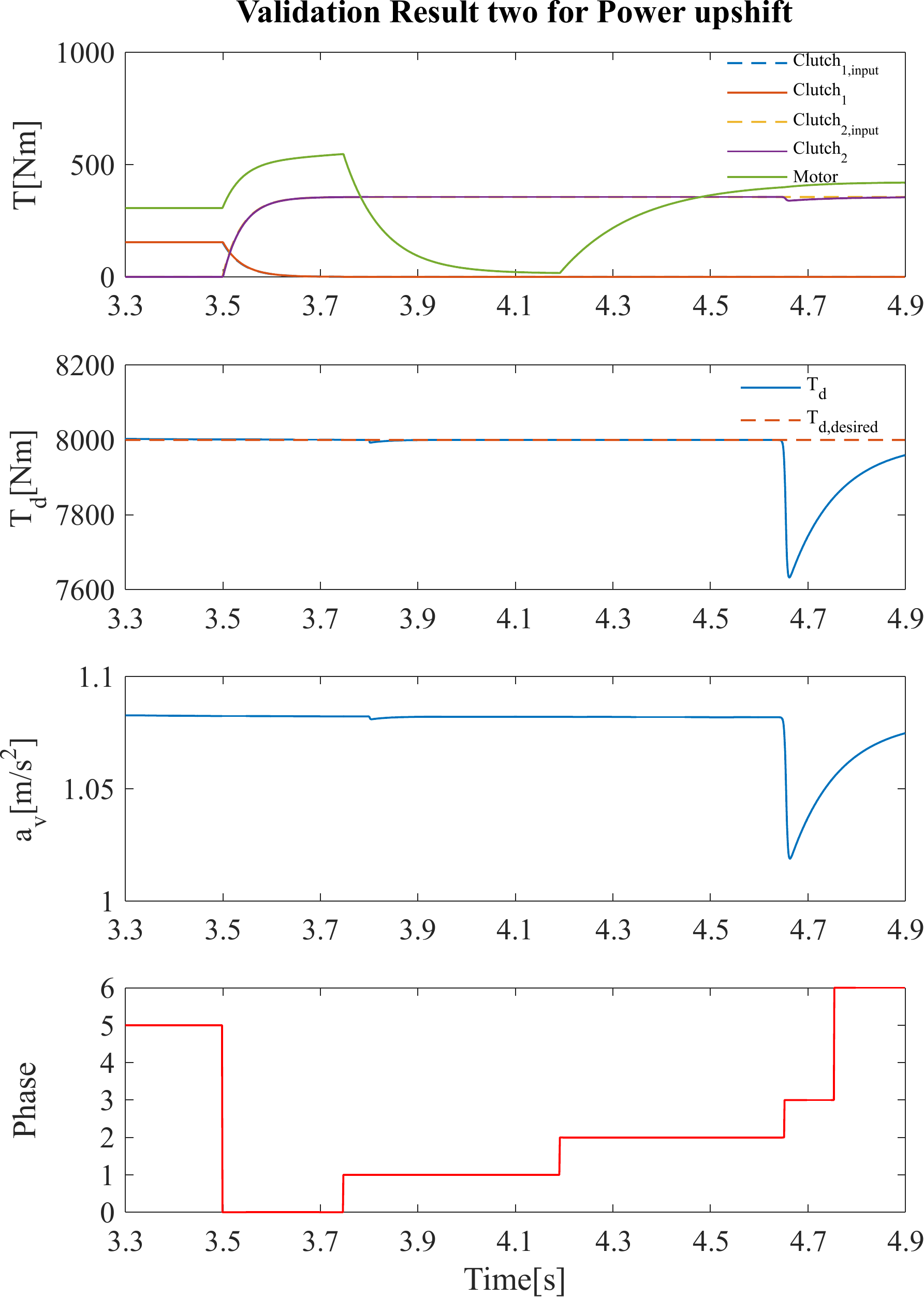}
	\caption{Control variables during the upshifting process}
	\label{fig:upshift_2}
\end{figure}

\section{Conclusion and outlook}

This article proposes a control strategy for the shifting process of construction machines
equipped with a dual-clutch transmission. We firstly conclude the main problems that occurred during the shifting process. First of all, although the initial intention of using the dual-clutch is to avoid the torque interruption during the shifting process, the torque output from the powertrain still has large fluctuations. Moreover, there is a dilemma between the rapidity and the smoothness of the shifting.

To solve the above problems, we divide the shifting process into different stages to break down the complex control task into sub-tasks and define a clear boundary line for each sub-phases to realize the modular control of the sub-process. By summarizing the characteristics of the sub-stages that appear in the downshift, we sum up the mathematical problems with unity from these sub-stages. The shift duration and the torque drop at the end of the shift are all affected by the slip acceleration. The first-order derivative of the slip acceleration is related to the changing rate of the motor and clutch torques. Therefore, the control of the shifting process is a control of the
changing rate of the torque according to the demand.

As shown in the simulation results, the control strategy by means of clutch and motor
torque control to achieve a smooth shifting process is feasible since the desired drive torque is well tracked. In addition, the highly dynamical actuator is not required to achieve the control performance, which shows the practical meaning of this control algorithm. When tracking the target drive torque during the shifting process, the controller firstly calculates the changing rate of the clutch torques; to meet the controllability condition
of the clutch torque, the controller calculates the corresponding
motor torque simultaneously. By applying this control algorithm, the systematically adjustable control performance is possible. To adjust the contradiction between rapidity and smoothness during the inertia phase, the controller uses two calibration parameters $\Gamma_{2,1}$ and $\Gamma_{2,2}$.

\subsection*{Outlook}
The limitation of the research is that the powertrain model used in this paper does not
consider the stiffness and damping of the transmission shaft. If the neglected damping and
stiffness are found to have a strong influence on the control effect in actual situations, the changing rate of the control signal is required to be constrained. The target is to avoid the coincide between the excitation signal and the characteristic frequency of the system, which may lead to a sympathetic vibration.

\section*{Acknowledgment}

We sincerely acknowledge: Dr. Steffen Mutschler, Robert Bosch GmbH, for his enthusiastic encouragement and insightful discussions on dual-clutch transmission; Hengping Zhao, Beijing Institute of Technology, for proof reading.


\bibliography{Literature.bib}{}
\bibliographystyle{IEEEtran}

\section*{Appendix}

\begin{table}[htbp]
\centering
\caption{Vehicle model parameters}
\begin{tabular}{cccc}
\hline
\hline
Symbol & Value & Unit & Description \\
\hline
$J_m$ & 1.5 & $kg m^2$ & Motor inertia\\
$m_v$ & 9450 & $kg$ & Vehicle mass\\
$r_{rad}$ & 615 & $mm$ & Wheel radius\\
$J_v$ & 3574.2 & $kg m^2$ & Vehicle equivalent inertia\\
$c_w$ & 0.8 & $-$ & Drag coefficient\\
$A_v$ & 0.8 & $m^2$ & Reference area\\
$\rho_{air}$ & 1.293 & $kg/m^3$ & Density of air\\
$i_1$ & 3.74 & $-$ & Gear one ratio\\
$i_2$ & 1.5 & $-$ & Gear two ratio\\
$i_{final}$ & 15.429 & $-$ & Final drive ratio\\
$J_m$ & 1.5 & $kg m^2$ & motor inertia\\
$\eta$ & 0.9 & $-$ & Final drive efficiency\\
$\theta_m$ & 0.08 & $s$ & Motor time constant\\
$\theta_c$ & 0.04 & $s$ & Clutch time constant\\
\hline
\hline
\end{tabular}
\label{vehicleparameter}
\end{table}

We give the parameters of the vehicle to help the readers to reproduce the results of our algorithm. It is based on the real parameters of the machine with acceptable measurement error. Based on our study, the algorithm is robust and still works in case the settings change. Notice that, for commercial use, you should ask for our permission since we have applied for the patent.

\end{document}